\documentclass[12pt,onecolume,a4paper]{article}
\usepackage{amssymb}
\usepackage{graphicx}

\begin{document}

% The title
{\Large The focus of light --\\linear polarization breaks the rotational symmetry of the focal spot}\\

% Authors, Affiliation etc.
{\large RALF DORN, SUSANNE QUABIS and GERD LEUCHS\\
 Zentrum f\"ur Moderne Optik, Universit\"at Erlangen--N\"urnberg,\\
 D--91058 Erlangen, Germany\\
 e--mails: dorn@kerr.physik.uni-erlangen.de; quabis@physik.uni-erlangen.de; leuchs@physik.uni-erlangen.de}\\

% Double-spaced from now on
%\setlength{\baselineskip}{24pt}

{\bf Abstract:}\\
We experimentally demonstrate for the first time that a linearly polarized beam is focussed to an asymmetric spot when using a high--numerical aperture focussing system. This asymmetry was predicted by Richards and Wolf [Proc.\,R.\,Soc.\,London A {\bfseries 253}, 358 (1959)] and can only be measured when a polarization insensitive sensor is placed in the focal region. We used a specially modified photodiode in a knife edge type set up to obtain highly resolved images of the total electric energy density distribution at the focus. The results are in good agreement with the predictions of a vectorial focussing theory.\\

\begin{figure}[h]
   \begin{center}
    	\scalebox{0.80}{\includegraphics{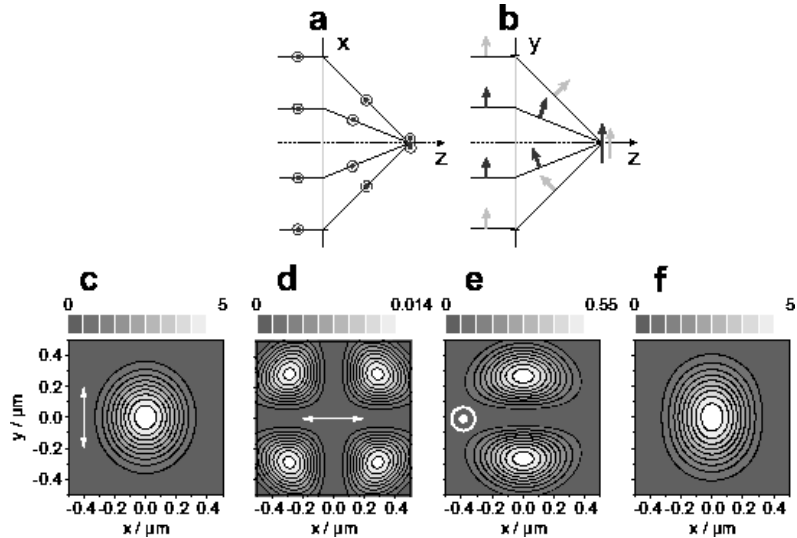}}\\
    	\caption{Plane wave model for focussing a beam which is linearly polarized parallel to the y--axis. Cross sections for two planes which contain the optical axis, (a) perpendicular and (b) parallel to the direction of polarization, show that the contribution to the total field of the rays propagating under a large angle to the optical axis is weaker in (b). Calculated intensity patterns for the three orthogonal field components (c--e) and total intensity distribution (f). The direction of polarization is indicated by arrows. (Calculations for NA=0.9, Gaussian input beam with a diameter (1/$e^2$) of 94\% of the diameter of the entrance pupil of the objective.
 	\label{abb:1}}
    \end{center}
\end{figure}

The ability to control and focus light is at the heart of the interdisciplinary field of optics and for many applications an exact knowledge of the structure of the focal field is required. 
Many areas in optical sciences make use of a tightly focussed light beam such as confocal microscopy \cite{TWilsonBuch} and optical data storage \cite{KinoSIL}. A highly concentrated and well matched field is also a necessary requirement for coupling to small quantum systems \cite{KimbleEnk} and applying light forces to microscopic particles \cite{NussenzveigOptPinzette}.
In the regime of strong focussing the widely used scalar theories are inadequate to describe the focal field. A vectorial focussing theory is required instead.
In the concrete example of a beam which is linearly polarized along the y--axis, the rays propagating in the xz- and in the yz-plane, respectively, contribute differently to the focal field (see figures \ref{abb:1}(a),(b)).
This has two closely related consequences.
The effective numerical aperture in the yz-plane is reduced and the focal spot is elongated in the direction of polarization of the input beam because the rays that propagate in this plane do not add up perfectly at the focus.
Associated with this elongation, field components in directions orthogonal to the direction of polarization of the input beam arise.
These effects are neglected in scalar descriptions \cite{SalesSmallestSpot}, which therefore do not predict the correct spot shape and size.

\noindent
All aspects of the field distribution of a tightly focussed linearly polarized light beam have been studied in theory taking into account the vector properties of the field \cite{RichardsWolf,Stamnes,KantGegenbauerPolynome,SheppardMultipoleExpansion}.
The theoretical calculations predict pronounced deviations from the results of the scalar theory. Nevertheless, a major part of the intensity\footnote{In the following, intensity always refers to electric energy density, which is the part of the field energy that couples to standard photodetectors and photosensitive materials.} is still predicted to be contained in a field component with an almost circular intensity distribution, which is polarized parallel to the direction of polarization of the incident beam (see figure \ref{abb:1} (c)). The remaining intensity appears in the two other mutually orthogonal field components. The orthogonal transverse component is calculated to show a four leaf clover shaped distribution (see figure \ref{abb:1} (d)) and does not contribute substantially to the shape of the focal field due to its minute intensity.
On the other hand, the intensity contained in a longitudinal z--component is calculated to contribute 16\% to the total intensity (for a numerical aperture of 0.9 and for homogeneous illumination). The calculated pattern of the intensity distribution of this component shows maxima away from the optical axis in the direction of the incident polarization (see figure \ref{abb:1} (e)). This causes the asymmetry of the total field (see figure \ref{abb:1} (f)).
When annular apertures are used for focussing, the relative strength of this longitudinal field and therefore the asymmetry is predicted to be enhanced \cite{SheppardAnnularApert}.

On the experimental side several studies of the focal intensity distribution have been reported. Using a CCD-camera Karman et al. \cite{KarmanCCDFokusMessung} achieved a large dynamic range of 5 orders of magnitude for the intensity. They focussed with a low numerical aperture, NA=0.01, for which polarization effects are too small to be detectable. The results were in excellent agreement with the predictions of the scalar theory. For the observation of polarization  effects it is not only important to focus the light with a high numerical aperture but also to detect light with an equally high effective numerical aperture. Firester et al. \cite{FiresterShengKnifeEdge} moved a knife edge through a strongly focussed beam (with numerical apertures up to 0.95). They observed only a small solid angle of the transmitted light because the detector was not in direct contact with the knife edge. Therefore they did not observe any asymmetry. Instead of the knife edge, one can also move a Ronchi ruling through the beam but in this case there is an additional requirement: one has to make a priori assumptions about the shape of the intensity distribution for the evaluation \cite{CohenApplOpt}. Collecting the light with a fibre tip typically provides the necessary high numerical detection aperture \cite{VaezIravani}. However, the metal coating around the tip introduces a significant polarization dependence \cite{RhodesNugentRobert}.
Another option is to scan small sub-wavelength beads through the focus and observe the scattered and reflected light \cite{WebbApplOpt,JuskaitisWilsonAmplitudePSF}. This technique, however, does also not readily disclose the focal intensity distribution in an unambiguous way. The detected signal depends on the emission characteristics of the bead \cite{WilsonOptComm} which in turn depends on the direction of the exciting field. This direction varies largely in the case of high numerical aperture focussing. If the detection solid angle is not exactly equal to 2$\pi$ or 4$\pi$, the overall detection efficiency will depend on the direction of the field exciting the bead. Therefore, one again has to make a priori assumptions about the measured distribution of the vectorial electric field {\bf E}.
Instead of isotropic beads, which do not have a preferred axis, one can also use individual oriented molecules the excitation of which depends sensitively on the direction of polarization of the incoming light. This property was used to determine the orientation of molecules on a surface \cite{HechtNovotnyPRL}. Single molecules oriented in three different mutually orthogonal directions could be used to scan the intensity distribution of all three polarization directions separately.
However, a quantitative comparison of the three patterns remains difficult as the correct relative scaling for the composition of the total intensity pattern requires the exact knowledge of the polarization--dependent sensitivity of the detection system.
An independent measurement which relates the position integrated intensities of different polarization components has been obtained only for the two transverse components \cite{HellDepolarisation}. To summarize the state of the art, individual aspects of the effects of strong focussing have been measured in several experiments, but none of them has measured the asymmetry of the polarization integrated total intensity distribution in the focal spot.

We present here the first direct experimental measurement of the total intensity distribution summing all three polarization components. We verify the elongated shape of the total intensity distribution due to the presence of all three mutually orthogonal field components. The results of our measurements are in agreement with the results of the vectorial focussing theory \cite{RichardsWolf,Stamnes}.

To measure the intensity distribution in the focus we use an experimental set up which is based on the knife-edge method. In this set up the transmitted light is detected in the immediate vicinity of the knife edge. Thus the effects of diffraction and propagation of the beam are minimized.
Our sensor is a specially designed p-i-n photodiode that is moved transversely in the focal plane using a piezo driven table operating in closed loop mode with a step width of 10nm to ensure high lateral resolution \cite{QuabisApplPhysB}. Parts of the light sensitive area are covered by a sharp edged opaque pad that plays the role of the knife edge. The photocurrent of the diode is proportional to the electric field intensity ($\propto |E|^2$) that is incident on the light sensitive area, which depends on the fraction of the beam blocked by the opaque structure. A more detailed description of the photodiode is given in the appendix. The data were taken with a laser intensity noise $<$0.6\% and the signal noise was reduced to $<$0.3\% by normalizing to the signal of the monitor diode. At each position of the edge the photocurrent is averaged for up to 100ms which further reduces the signal noise to $<$0.1\%. For every scan of the edge the signal is corrected for offsets when the beam is completely blocked by the opaque structure and is normalized to one when the whole beam hits the light sensitive area. The differentiation of the signal with respect to the detector position yields the line integrals of the intensity distribution in one transverse direction, parallel to the edge. This is equivalent to the projection of the 2D intensity distribution on to a line perpendicular to the edge. The curves are differentiated with the Savitzky-Golay algorithm \cite{NumericalRecipesInC} using five adjacent points.\\

\begin{figure}[h]
    \begin{center}
       \scalebox{0.80}{\includegraphics{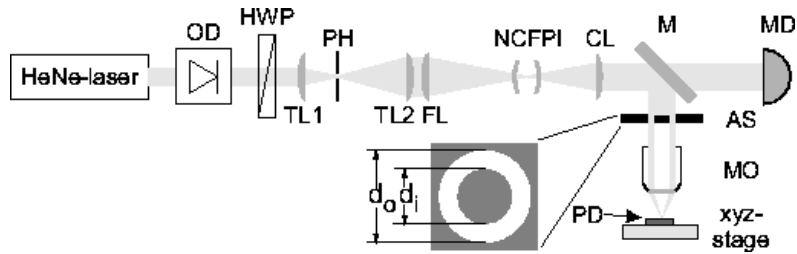}}\\
       \caption{Experimental set up to measure the focal intensity distribution. MO: Leica plan--apo 100x, 0.9, OD: optical diode, HWP: half--wave plate, TL: telescope lenses, PH: pinhole, FL: focussing lens, NCFPI: non--confocal Fabry--Perot interferometer, CL: collimating lens, AS: aperture stop, M: four mirrors for polarization insensitive deflection (details not shown), MD: monitor diode, PD: photo diode partially covered with gold--zinc alloy
    	\label{abb:2}}
    \end{center}
\end{figure}

\noindent To provide a linearly polarized and well defined input field, a helium neon laser beam ($\lambda$=632.8nm) is sent through a non--confocal Fabry--Perot interferometer operating as a mode cleaner. The transmitted beam is collimated to a TEM$_{00}$-mode with a beam radius of $\omega_G$=1.7mm (see figure \ref{abb:2}).
Using a half-wave plate, the direction of polarization was rotated with respect to the edge. 
There was no significant beam displacement due to the rotation of the half-wave plate, documented by the fact that the maximum displacement of the peak positions in the sensor plane is below 20nm.
Measurements for 18 different orientations of the polarization relative to the edge were performed to reconstruct the intensity distribution in the focal plane using the Radon back transformation \cite{Rowland}. For measurements with an annular aperture, a stop is placed directly in front of the microscope objective to block the inner part of the input beam. The stop is a glass substrate coated with an opaque disc (d$_i$=2.0mm). The diameter of the entrance pupil of the microscope objective is 3.6mm.

\begin{figure}[h]
    \begin{center}
   	\scalebox{0.80}{\includegraphics{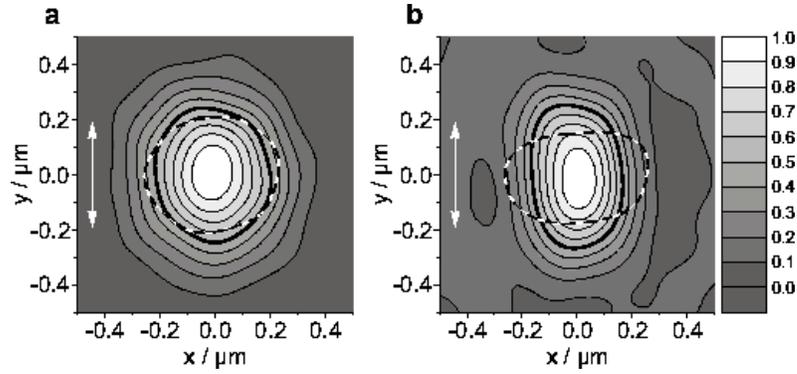}}\\
     	\caption{(a)Reconstructed intensity distribution in the focal plane for a linearly polarized input field without annular apperture. The contour line at half the maximum value (thick black line) is shown rotated by 90$^\circ$ (dashed line). The spot is elongated in the direction of polarization (indicated by the arrow). Using an annular aperture (b) the elongation increases.
        \label{abb:3}}
    \end{center}
\end{figure}

In figure \ref{abb:3}(a) and (b) we present distributions of the focal intensity for a linearly polarized beam that was focussed without and with annular apertures respectively. The distributions were reconstructed from the experimental data and both show the elongation of the focus.
It is clearly visible that the elongation of the focus increases when an annular aperture is inserted.
Figures \ref{abb:3} (a) and (b) show cross sections through the reconstructed distributions.
The focal spot size is reduced from 0.40$\lambda^2$ (theory: 0.37$\lambda^2$) without annular aperture to 0.35$\lambda^2$ (theory: 0.28$\lambda^2$) with annular aperture. This is mainly due to a reduction of the width of the profile in the direction perpendicular to the  initial polarization. In this direction the increase of the sidelobes due to the annular aperture is most pronounced.\\

\begin{figure}[h]
    \begin{center}
       \scalebox{0.80}{\includegraphics{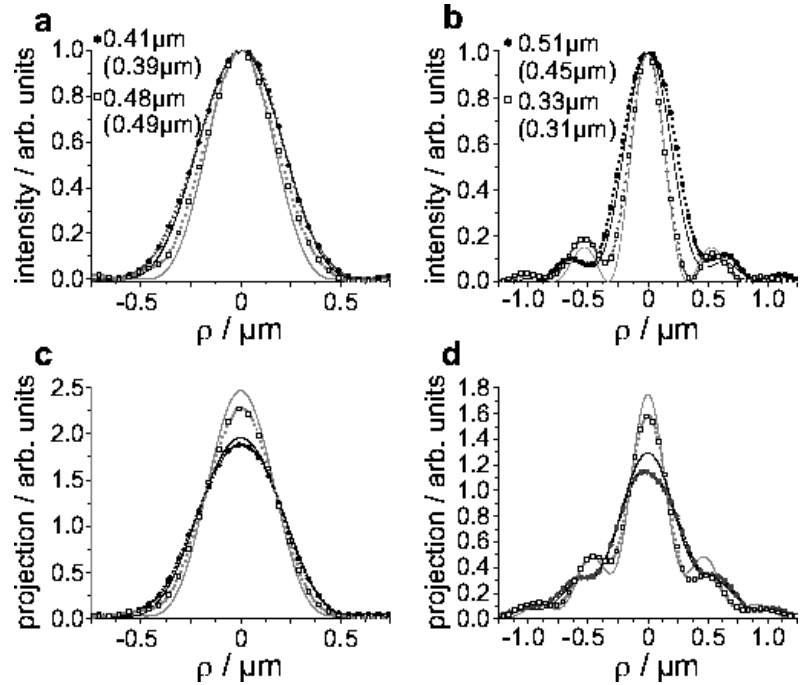}}\\
       \caption{Cross sections through the reconstructed intensity distributions (a,b) along a line parallel (solid circles) and perpendicular (open squares) to the direction of polarization.
Values for the measured FWHM are provided together with theoretical values in brackets.
(c,d) Measured projections of the intensity distribution on to a line parallel (solid circles) and perpendicular (open squares) to the direction of polarization. (a,c) without and (b,d) with annular aperture. Solid lines: theoretical curves
       \label{abb:4}}
    \end{center}
\end{figure}

\noindent The asymmetry is even more obvious if one goes back to the input data for the tomographic reconstruction i.e. the initial line integrals of the intensity distribution (see figures \ref{abb:4} (c)and (d)).
Parallel to the direction of polarization the distribution is broader than in the orthogonal direction.

It is remarkable that the measured and the theoretical profiles are in good agreement even in the shapes of the slopes.
The measured height ratios of these projections of 0.83 (theory: 0.79) without and 0.73 (theory: 0.74) with annular aperture are close to the theoretical values.\\

\begin{figure}[h]
    \begin{center}
    \scalebox{0.80}{\includegraphics{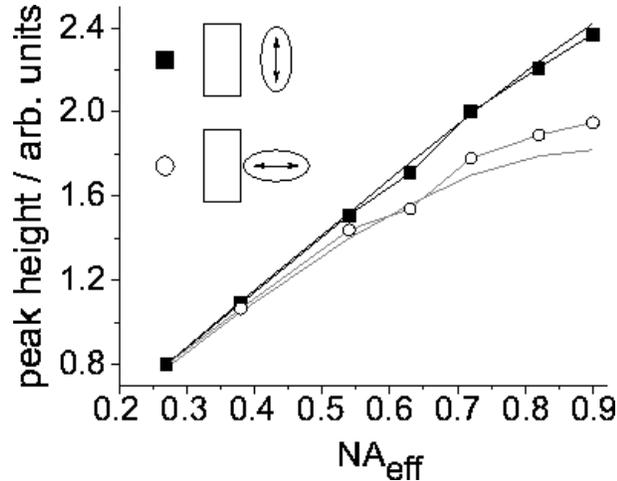}}\\
    \caption{Peak height of the projected intensity distributions vs. effective numerical aperture. Solid squares: projections on to a line perpendicular to the direction of polarization, open circles: projections on to a line parallel to the direction of polarization along with theoretical curves (solid lines). An input field with a homogeneous intensity distribution was used. Note that the polarization anisotropy appears only for high numerical apertures.
    \label{abb:5}}
    \end{center}
\end{figure}

\noindent Also note, that we carefully checked that the elongation of the focus is not an artefact caused by any residual polarization dependent interaction with the edge. The experimental evidence is that the elongation completely vanishes when the effective numerical aperture of the microscope objective (Leica, Plan Apo, 100x, 0.9) is reduced stepwise to 0.27 by restricting the diameter d$_{\rm o}$ of the input beam by using an aperture stop (figure\,\ref{abb:5}).

The observed deviation between theory and experiment, 5\% for the measured data and up to 20\% for the deduced focal area, may be attributed to residual imperfections of the set-up each causing deviations at the percentage level. Firstly, the Debye-approximation used to obtain the theoretical curves neglects diffraction at the aperture rim. Second, the objective is man made; in addition to radially dependent Fresnel-losses, there may be stress-induced birefringence and deviations from the perfect surface geometry. Third, similar arguments hold for the overall beam geometry and the detector. The positioning precision e.g. of the annular aperture with respect to the objective is about 50$\mu$m. Fourth, the Radon back transformation requires perfectly centered data. A centering error of 10$^{-3}$ causes already changes of about 10$^{-2}$ for the spot area. Fifth, the differentiation of the signal is sensitive to beam instabilities. The mechanical stability of the set up has been improved until the position fluctuation of the focal spot was below 1nm. We estimate that the effect of all these imperfections of the set up and of the theory can well explain the observed discrepancy.

\begin{figure}[h]
    \begin{center}
    \scalebox{0.80}{\includegraphics{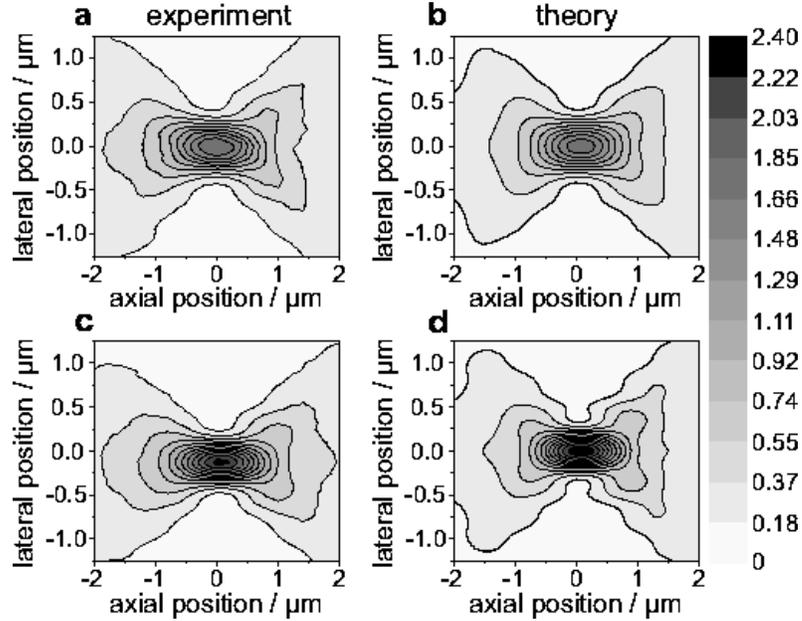}}\\
    \caption{Projections of the intensity distribution for a linearly polarized input beam. Measured (a,c) and calculated (b,d) projections on to planes which contain the optical axis which are parallel (a,b) and perpendicular (c,d) to the direction of linear polarization. A spherical aberration (0.0075$\lambda$ rms) of the wavefront was assumed in the calculations.
    \label{abb:6}}
    \end{center}
\end{figure}

Line integrals of the intensity distribution were measured also for different defocus positions (see figure \ref{abb:6}). This yields a cross section of the propagating beam containing the optical axis. The observed asymmetry with respect to the focal plane is caused by minute deviations of the wavefront \cite{SheppardAberrations} introduced by the microscope objective which are at the limit of a state-of-the-art production process. For the comparison of the experimental results with numerical calculations we considered spherical aberrations and found the best agreement for a wavefront with an rms error of 0.0075$\lambda$, which is in excellent agreement with the manufacturer`s specifications \cite{VollrathPrivComm} conveyed to us after the measurement. Note, that the projections in figure 6 are distinctly different for the two planes parallel and perpendicular to the initial polarization. Also note, that the effect of these aberrations was observed in the defocus series. According to theory it does not influence the focal size and shape in a measurable way.\\

\noindent In summary, we report the first reconstruction of the elongated focus caused by polarization effects. The set up is also sensitive enough to detect quantitatively the effect of the minute spherical aberrations introduced by the focussing microscope objective.
The longitudinal field component showing up in the focal plane may help to improve the spectroscopic investigation of the 3D orientation of a single molecule near a surface \cite{MacklinScience,EmpedoclesNature} and evidence for the existence of the longitudinal field component has already been reported \cite{HechtNovotnyPRL}.
The good agreement between the measured and calculated intensity profiles we report here allows one also to quantitatively infer the appearance of a longitudinal field component.\\

{\bf Acknowledgements:}\\
We are grateful to W.\,Vollrath from Leica Microsystems for his support. We thank G.\,D\"ohler for the fabrication of the p-i-n--diode. We appreciate helpful discussions with Z.\,Hradil, H.\,J.\,Kimble and P.\,Leiderer.

\newpage
{\bf Appendix:}\\
The photodiode used in our experiment is a p-i-n AlGaAs diode that consists of a GaAs substrate, an n-doped layer of AlGaAs, a layer of intrinsic GaAs, a p-doped layer of AlGaAs and a p-doped GaAs cover layer (seen from the substrate to the surface). The cover layer 10nm thick is necessary to protect the AlGaAs from corrosion. During the fabrication process, care must be taken not to damage this layer. For a diode that showed scratches that were up to 30nm deep as measured by an AFM, the photocurrent showed unacceptably large fluctuations when the focussed spot was scanned over the active surface.
The diode used in the experiment showed a surface roughness of below 3nm rms and the corresponding fluctuations were below 0.1\% (peak-to-valley). The active surface of the device is 200$\mu$m $\times$ 260$\mu$m wide and is surrounded by rectangular contact pads to tap the photocurrent which each have a size of 200$\mu$m $\times$ 120$\mu$m.

\begin{figure}[h]
    \begin{center}
      \scalebox{0.75}{\includegraphics{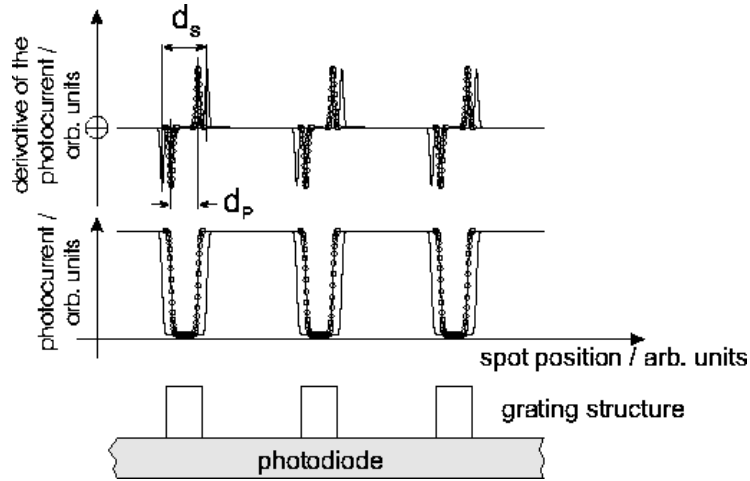}}\\
      \caption{Schematic distribution of the photo current (b) and its derivative (a) with respect to the position of the focal spot relative to a grating structure (c) placed on the photodiode.
The distance between the extrema before and behind one bar of the grating in (a) depends on the direction of polarization of the incident beam relative to the grating. d$_S$ denotes this distance for a polarization perpendicular (solid line) to the bars and d$_P$ the distance for a polarization parallel ($\circ$) to the bars.
      \label{abb:7}}
    \end{center}
\end{figure} 

\noindent The opaque material that forms the knife edge plays an important role. To allow for a thin layer the opaque structure should show a small transmission and therefore preferably be made of metal. However, for a metallic edge material, one would typically expect a polarization dependent interaction with the light field. For several edge materials we observed that the position of the strongest increase of the photocurrent depends on the direction of the polarization of the input field with respect to the edge when scanning the edge relative to the focus.
To characterize this polarization dependent shift we determined the "duty-cycle" of the signal measured for different grating structures on the active surface. Each grating was measured twice once with the polarization of the input field being parallel and perpendicular to the grating. The difference in the "duty cycle" for the two directions of polarization of the input field allows one to infer the relative lateral shift of the two transverse field components in the focal plane (see figure \ref{abb:7}). This procedure has the advantage that it does not require any absolute position measurements.

\begin{figure}[h]
    \begin{center}
     \scalebox{0.50}{\includegraphics{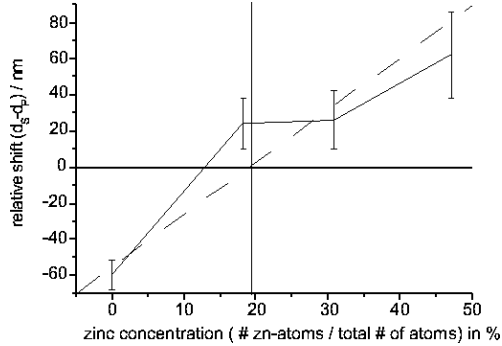}}\\
     \caption{Measured difference (d${_S}-$d$_P$) of the interval between the extrema in the line integrated projections. This difference is caused by the polarization dependent interaction with the edge as a function of the zinc concentration. 
A least square fit (dashed line) was used to find the optimum zinc concentration (about 20\%).
     \label{abb:8}}
    \end{center}
\end{figure}

The shift was determined for different edge materials, gold and different gold-zinc alloys. It turned out that this shift depends on the zinc-concentration of the edge material (see figure \ref{abb:8}).

\noindent We found no significant relative shift for a zinc concentration of about 20\%. Therefore, all measurements presented in the main part of the paper were performed using the edge material with the optimized concentration and a thickness of about $\lesssim$ 200nm, resulting in a transmission below 1\%.
For the chosen alloy a good agreement with the results predicted by theory was observed. All other materials which were tested so far were excluded as well behaved edge materials.
It is worth noting that for the well behaved edge that caused a minimal differential shift between the x and y polarization component (see figure 8) we also did not observe any significant shift for the z-component. Here x and y refer to the two transverse components and z to the longitudinal polarization component. x is parallel to the edge whereas y and z are perpendicular to the edge.
The mechanism that for some materials leads to a displacement of individual field components is not yet completely clarified. A possible mechanism under study is the excitation of surface plasmons. However, we want to stress that the above mentioned test gives phenomenological evidence that the diode with the well behaved edge material is suitable to perform the described measurements of the asymmetry of the focus.

\end{document}